\documentclass[prb,twocolumn,showpacs]{revtex4}
\usepackage{amsmath}
\usepackage{dcolumn}
\usepackage{graphicx}
\usepackage{bm}

\begin{document}

\title{Topologically distinct classes of valence bond solid states with
their parent Hamiltonians}
\author{Hong-Hao Tu and Guang-Ming Zhang}
\email{gmzhang@mail.tsinghua.edu.cn}
\affiliation{Department of Physics, Tsinghua University, Beijing 100084, China}
\author{Tao Xiang}
\affiliation{Institute of Physics, Chinese Academy of Sciences, P.O.Box 603, Beijing
100190, China;\\
Institute of Theoretical Physics, Chinese Academy of Sciences, P.O.Box 2735,
Beijing 100190, China}
\author{Zheng-Xin Liu and Tai-Kai Ng}
\affiliation{Department of Physics, Hong Kong University of Science and Technology,
Kawloon, Hong Kong, China}
\date{\today }

\begin{abstract}
We introduce a general method to construct one-dimensional translationally
invariant valence bond solid states with a built-in Lie group $G$ and derive
their matrix product representations. The general strategies to find their
parent Hamiltonians are provided so that the valence bond solid states are
their unique ground states. For quantum integer spin-$S$ chains, we discuss
two topologically distinct classes of valence bond solid states: One
consists of two virtual $SU(2)$ spin-$J$ variables in each site and another
is formed by using two $SO(2S+1)$ spinors. Among them, a new spin-$1$
fermionic valence bond solid state, its parent Hamiltonian, and its
properties are discussed in detail. Moreover, two types of valence bond
solid states with $SO(5)$ symmetry are further generalized and their
respective properties are analyzed as well.
\end{abstract}

\maketitle

\section{Introduction}

In recent years, the study of topological order has become a common issue in
condensed matter physics and quantum information theory. \cite%
{Wen-2004,Kitaev-2006,Feng-2007,Oshikawa-2006,Bombin-2007,Aguado-2008,Cirac-2008}
Historically, this concept was proposed to describe fractional quantum Hall
states, \cite{Wen-1990} which are incompressible quantum liquids with a
finite energy gap to all bulk excitations. These new quantum phases of
matter can not be described by a local order parameter with spontaneous
symmetry breaking. So the discovery of fractional quantum Hall effect brings
great challenge to the Ginzburg-Landau theory, which is a corner-stone
paradigm to characterize phases and phase transitions in condensed matter
physics. From the viewpoint of quantum information theory, the appearance of
a long-range quantum entanglement plays an essential role in the
topologically ordered states. However, a general multipartite-entanglement
measure that captures the most relevant physical properties is still
lacking, because the number of parameters required to describe a quantum
many-body state usually grows exponentially with the particle number.

The topological order appears not only in the two-dimensional\ fractional
quantum Hall states but also in one-dimensional systems, for instance, the
quantum integer-spin chains. In 1983, Haldane predicted that quantum
integer-spin antiferromagnetic Heisenberg chains have an exotic energy gap. %
\cite{Haldane-1983}\ Later, Affleck, Kennedy, Lieb, and Tasaki (AKLT) \cite%
{Affleck-1987}\ found a family of exactly solvable integer-spin models with
valence bond solid (VBS) ground states, and the presence of an excitation
gap can be\ proved rigorously. In a spin coherent state representation,
these VBS states share a striking analogy to the fractional quantum Hall
states. \cite{Arovas-1988}\ Although the two-point spin correlation decays
exponentially, den Nijs and Rommelse \cite{den Nijs-1989}\ found a hidden
topological order in the $S=1$ VBS state, which is characterized by
non-local string order parameters. For the $S=1$ VBS state, the long-range
string order and the fourfold degeneracy in an open chain can be understood
as natural consequences of a hidden $Z_{2}\times Z_{2}$\ symmetry breaking. %
\cite{Kennedy-1991,Oshikawa-1992} For the standard integer-spin Heisenberg
models, the existence of spin-$S/2$ edge states was also verified by quantum
field-theory mappings \cite{TKNg-1994}\ and numerical calculations \cite%
{SJQin-1995}, which coincide with the VBS picture of the AKLT models.\
Experimentally, the VBS\ picture for $S=1$ Haldane chain was supported by
the electron spin resonance studies \cite{Hagiwara-1990,Glarum-1991}\ of the
compound Ni(C$_{2}$H$_{8}$N$_{2}$)$_{2}$NO$_{2}$(ClO$_{4}$) (NENP), the NMR
imagining \cite{Tedoldi-1999}\ and the magnetic neutron scattering study %
\cite{Xu-2000}\ of the quasi one-dimensional material Y$_{2}$BaNiO$_{5}$.

The one-dimensional VBS states can be represented in a matrix-product form. %
\cite{Fannes-1989,Klumper-1991,Suzuki-1995} Moreover, it was found \cite%
{Ostlund-1995} that density matrix renormalization group \cite{White-1992}
(DMRG), the most powerful numerical method for one-dimensional quantum
systems, converges to a matrix-product wave function as its fixed point.
This important observation stimulates the formulation of the numerical
techniques in one dimension by using the matrix-product variational wave
functions. \cite{Vidal-2004} From a quantum information perspective, the
validity of the matrix-product variational ansatz depends on whether the
true quantum ground states of the system obey an area law of entanglement
entropy. \cite{Eisert-2008} In this sense, the VBS states are only slightly
entangled because their entanglement entropies have upper bounds even in the
thermodynamic limit. Recently, these VBS states have received considerable
attentions since they provide a playground to test the proposed measures of
multipartite entanglement. \cite{Verstraete-2004,HFan-2004,Hatsugai-2007}
Towards the potential applications, it was suggested that the VBS states
ensure measurement-based quantum computation. \cite{Verstraete-Cirac-2004}
For the $S=1$ VBS state of the AKLT model, Brennen and Miyake \cite%
{Brennen-2008} have shown that a gap-protected measurement-based quantum
computation can be performed within the degenerate ground states.

In this paper, we will introduce one-dimensional translationally invariant
VBS states with a built-in Lie group $G$ and present a general method to
construct their parent Hamiltonians. For quantum integer spin-$S$ chains, we
focus on two classes of VBS states. The local spin-$S$ states are formed by
two virtual $SU(2)$ spin-$J$ variables in the first class and by two $%
SO(2S+1)$ spinors in\ the second one. To illustrate the method, we choose
the $S=1$ fermionic VBS state with virtual spin $J=1$ as an explicit example
to find the parent Hamiltonian. Apart from the $S=1$ VBS state with virtual
spin $J=1/2$ and the $S=2$ VBS state with virtual spin $J=3/2$ as the
intersection elements, the VBS states of the two classes are shown to be
topologically distinct to each other, which can be characterized by their
edge spin representations in open chain systems. We also apply our method to
investigate $SO(5)$ symmetric spin chains and discuss several VBS states
with interesting properties.

The outline of this paper is structured as follows. In Sec. II, we will
introduce VBS states with a Lie group symmetry and derive their matrix
product representations. In Sec. III, we will focus on quantum integer-spin
chains and study two topologically distinct classes of VBS states, including
spin-$S$ VBS states formed by virtual $SU(2)$ spin-$J$ particles and by
virtual $SO(2S+1)$ spinor particles. Moreover, a spin-$1$ fermionic VBS
state is extensively studied as an example and we construct its parent
Hamiltonian explicitly. Sec. IV is devoted to the $SO(5)$ symmetric VBS
states and their physical properties. In Sec. V, some conclusions are drawn.

\section{General construction of VBS states}

\subsection{Matrix-product form}

\begin{figure}[tbp]
\centering \includegraphics[scale=1]{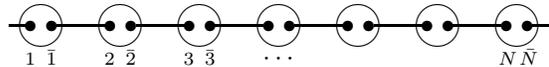}
\caption{The schematic of the VBS states with a built-in Lie group $G$. Each
dot denotes a virtual particle transforming under $G_{D}$ irreducible
representations of Lie group $G$. The solid lines represent valence-bond
singlets formed by two virtual $G_{D}$ irreducible representations on the
neighboring sites, and the circles denote the projections of the virtual
particles in each lattice site onto the physical $G_{d}$ irreducible
representations. }
\label{Fig:VBS}
\end{figure}

We begin with a quantum one-dimensional chain with $N$ lattice sites. In
each site, the states $\left\{ \left\vert m\right\rangle \right\} $ $%
(m=1,\ldots ,d)$ transform under a $d$-dimensional irreducible
representation (IR)$\ G_{d}$ of a Lie group $G$. Let us imagine that the
physical Hilbert space is formed by two virtual \textit{identical}
particles, whose internal quantum numbers $\left\{ \left\vert \alpha
\right\rangle \right\} $ $(\alpha =1,\ldots ,D)$ transform under the $D$%
-dimensional IR $G_{D}$ of the same Lie group $G$. Here we require that both
singlet representation $G_{I}$ and IR $G_{d}$ are included in the tensor
product decomposition of two $G_{D}$'s. The first requirement means that $%
G_{D}$ is a \textit{self-conjugate} IR, i.e., the complex conjugate
representation of $G_{D}$ is equivalent to $G_{D}$. The latter requirement
can be implemented by the projection operator onto the physical Hilbert
space \cite{Cirac-2007}
\begin{equation}
P=\sum_{m=1}^{d}\sum_{\alpha ,\beta =1}^{D}P_{\alpha ,\beta
}^{[m]}\left\vert m\right\rangle \left\langle \alpha ,\beta \right\vert ,
\label{eq:projection}
\end{equation}%
where $P_{\alpha ,\beta }^{[m]}$ is the Clebsch-Gordan coefficient defined
by $P_{\alpha ,\beta }^{[m]}=\langle G_{d},m|G_{D},\alpha ;G_{D},\beta
\rangle $. For VBS states in a periodic chain, each lattice site forms a
valence-bond singlet $\left\vert I\right\rangle $ with its neighboring sites
by pairing two virtual particles (See Fig. \ref{Fig:VBS}). Thus, the wave
functions of the VBS states can be expressed as%
\begin{equation}
\left\vert \Psi \right\rangle =(\otimes _{k=1}^{N}P_{k,\bar{k}})\left\vert
I\right\rangle _{\bar{1}2}\left\vert I\right\rangle _{\bar{2}3}\cdots
\left\vert I\right\rangle _{\bar{N}1},  \label{eq:VBSstate}
\end{equation}%
where the valence-bond singlet $\left\vert I\right\rangle $\ is given by%
\begin{equation}
\left\vert I\right\rangle _{ij}=\sum_{\alpha ,\beta =1}^{D}Q_{\alpha ,\beta
}\left\vert \alpha \right\rangle _{i}\otimes \left\vert \beta \right\rangle
_{j}.  \label{eq:singlet}
\end{equation}%
Here $Q_{\alpha ,\beta }=\langle G_{I},I|G_{D},\alpha ;G_{D},\beta \rangle $
is the Clebsch-Gordan coefficient of contracting two virtual $G_{D}$
representations to form a singlet representation $G_{I}$.

In the present formalism, the VBS states can be easily written in a
matrix-product form. Since $P_{\alpha ,\beta }^{[m]}$ and $Q_{\alpha ,\beta
} $ can be viewed as the matrix elements of $D\times D$ matrices of $P^{[m]}$
and $Q$, the VBS states in Eq. (\ref{eq:VBSstate}) can be thus written as
the following matrix-product form:%
\begin{equation}
\left\vert \Psi \right\rangle =\sum_{m_{1}\cdots m_{N}}\mathrm{Tr}%
(A^{[m_{1}]}A^{[m_{2}]}\cdots A^{[m_{N}]})\left\vert m_{1}\cdots
m_{N}\right\rangle ,  \label{eq:MPSrep}
\end{equation}%
where $A^{[m]}=P^{[m]}Q$ is a $D\times D$ matrix.

In periodic boundary conditions, the VBS states are invariant under lattice
translation and transformation of the Lie group $G$ by construction.
Although no local order parameters can be found to characterize these
states, the $A^{[m]}$ matrices can fully determine their physical properties
and render a \textquotedblleft local\textquotedblright\ description. In open
boundary conditions, edge states emerge at the two ends of the chain and
then the matrix-product form of the VBS states is given by%
\begin{equation}
\left\vert \Psi _{\alpha ,\beta }\right\rangle =\sum_{m_{1}\cdots
m_{N}}(A^{[m_{1}]}A^{[m_{2}]}\cdots A^{[m_{N}]})_{\alpha ,\beta }\left\vert
m_{1}\cdots m_{N}\right\rangle ,  \label{eq:OBC-MPS}
\end{equation}%
where the matrix indices $\alpha ,\beta $ denote the edge states. These edge
degrees of freedom are described by two fractionalized particles
transforming under $G_{D}$ representation of the Lie group $G$. Actually,
the edge states and their representations are characteristic features of the
VBS states, because they fully determine the local $A^{[m]}$ matrices.

Another useful way to represent the VBS states is to use boson or fermion
realization methods. To illustrate this method, we trace back to the tensor
product decomposition of IRs of Lie algebras. According to the group theory,
the physical states $\left\vert m\right\rangle $ under the exchange of the
two identical virtual particles is either symmetric or antisymmetric,
depending on $G_{d}$ and $G_{D}$. Thus, the two virtual particles with
fermion statistics create the antisymmetric states, while the bosonic
particles yield the symmetric ones. In some cases, there are still several
channels with the same exchange symmetry and additional projection has to be
used to single out the physical states in $G_{d}$. For example, the
fermionic realization of a spin-$2$ VBS state with virtual spin $J=3/2$ was
considered in Ref. \cite{Tu-2008}. In this case, both site-quintet states ($%
S=2$) and site-singlet state ($S=0$) are allowed for two spin-$3/2$ fermions
on a single site and an extra projection can remove the unphysical
site-singlet state. There also exists the Schwinger boson realization which
symmetrizes several fundamental IRs to form higher dimensional $G_{d}$'s. %
\cite{Arovas-1988,Greiter-2007,Arovas-2008,Rachel-2008}\ In fact, all these
boson/fermion realization methods play the role of (sometimes partially)
deleting the unphysical states.

\subsection{Parent Hamiltonian: Locating the null space}

Following the spirit of the AKLT model, one can construct the parent
Hamiltonians such that the VBS states in Eq. (\ref{eq:VBSstate})\ are their
unique ground states. It is most convenient to work with the matrix-product
form. For the matrix product states in Eq. (\ref{eq:MPSrep}), one can
readily find that their reduced density matrix $\rho _{l}$ of $l$ successive
sites has a rank of $D^{2}$ at most. This suggests that the reduced density
matrix $\rho _{l}$ of these VBS\ states always have null space for
sufficient large $l$. These states are annihilated by the local projection
operators supported in the null space. Hence, they are always exact
zero-energy ground states of the translationally invariant Hamiltonians%
\begin{equation}
H=\sum_{i}h_{i},
\end{equation}%
where $h_{i}$ contains a sum of the positive semi-definite
projection operators supported in the null space from site $i$ to
$i+l-1$. Previously, the parent Hamiltonians of the matrix product
states for spin-ladder systems had been studied by similar
methods.\cite{km}
\begin{figure}[tbp]
\centering \includegraphics[scale=1]{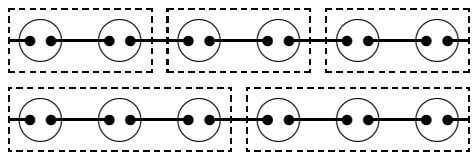}
\caption{The schematic of the \textquotedblleft
coarse-graining\textquotedblright\ process that converts the spins of
successive sites to block spins. This procedure leads to a matrix product
state with block spins, and the null space of a block-spin reduced density
matrix can be identified.}
\label{Fig:BlockedVBS}
\end{figure}

Let us begin with the simplest cases with only nearest-neighbor
interactions. Now the null space can be obtained from the VBS picture of
these states. The Hilbert space of two neighboring sites can be divided into
a direct sum of different IR channels according to the tensor product
decomposition $G_{d}\otimes G_{d}$. Once a valence-bond singlet of two
virtual $G_{D}$'s is formed, the remaining two particles of adjacent sites
can transform under a direct sum of IRs resulting from $G_{D}\otimes G_{D}$.
Therefore, the IR\ channels contained in $G_{d}\otimes G_{d}$ but absent in $%
G_{D}\otimes G_{D}$ constitute the null space in the $2$-site reduced
density matrix.

The above steps to locate the null space can be embedded in a matrix-product
formalism. Practically, we rewrite the matrix-product states in Eq. (\ref%
{eq:MPSrep}) as%
\begin{equation}
\left\vert \Psi \right\rangle =\mathrm{Tr}(g_{1}g_{2}\cdots g_{N}),
\end{equation}%
where the local matrix $g_{i}$ is defined by%
\begin{equation}
g_{i}\equiv \sum_{m_{i}}A^{[m_{i}]}\left\vert m_{i}\right\rangle .
\end{equation}%
To locate the null space, we resort to a \textquotedblleft
coarse-graining\textquotedblright\ procedure which converts the spins of
adjacent sites to block spins. Since the VBS states are invariant under
lattice translation, we can block the spins in sites $1$ and $2$ as%
\begin{eqnarray}
g_{1}g_{2} &=&\sum_{m_{1},m_{2}}A^{[m_{1}]}A^{[m_{2}]}\left\vert
m_{1},m_{2}\right\rangle  \notag \\
&=&\sum_{G_{12},M_{12}^{G}}B^{[G_{12},M_{12}^{G}]}\left\vert
G_{12},M_{12}^{G}\right\rangle ,
\end{eqnarray}%
where the $D\times D$ matrices $B^{[G_{12},M_{12}^{G}]}$ are given by
\begin{equation}
B^{[G_{12},M_{12}^{G}]}=\sum_{m_{1},m_{2}}A^{[m_{1}]}A^{[m_{2}]}\langle
G_{12},M_{12}^{G}|G_{d},m_{1};G_{d},m_{2}\rangle .
\end{equation}%
Here $G_{12}$'s distinguish the IRs of nearest-neighbor bond spin channels
and $\left\vert G_{12},M_{12}^{G}\right\rangle $ are the states in IR
channel $G_{12}$. Correspondingly, $\langle
G_{12},M_{12}^{G}|G_{d},m_{1};G_{d},m_{2}\rangle $ is the Clebsch-Gordan
coefficient to combine the states of $G_{d}$'s into the states of $G_{12}$.
In the example of $SU(2)$, $G_{12}$ denotes for the total bond spin $S_{12}$
and $-S_{12}\leq M_{12}^{S}\leq S_{12}$. After this \textquotedblleft
coarse-graining\textquotedblright\ procedure, the VBS states are transformed
to a matrix-product form with $2$-site block spins, characterized by the
block-independent matrices $B^{[G,M^{G}]}$. Since the $2$-site block spin
states $\left\vert G_{12},M_{12}^{G}\right\rangle $ form a complete
orthogonal set, the null space in the reduced density matrix of a $2$-site
block is given by those IR channels with $B^{[G_{12},M_{12}^{G}]}=0$.

The null space for more than two adjacent sites is no longer easily\
visualized. However, the blocking process of $g$ matrices can be proceeded
without any fundamental difficulties (See Fig. \ref{Fig:BlockedVBS}). In
Sec. III B, we will study the spin-$1$ fermionic VBS state by using this
powerful method.

The uniqueness of the VBS ground states of the constructed Hamiltonians has
to be further clarified. Generally, the ground state degeneracy will occur
if there exists another state with a larger null space in the reduced
density matrix of the present interaction range. To lift the degeneracy, one
can locate the null space in an extended range by blocking more spins and
modify the Hamiltonian correspondingly. To justify the uniqueness, it is
helpful to numerically diagonalize an open chain Hamiltonian with several
lattice sites. If the numerically calculated ground-state degeneracy is $%
D^{2}$, i.e., the ground states are all contained in the matrix-product
form, one can prove the uniqueness of VBS ground states by a mathematical
induction method. The basic idea is to assume that the VBS states $\left|
\Psi _{\alpha ,\beta }\right\rangle $ in Eq. (\ref{eq:OBC-MPS}) are the only
ground states of a projector Hamiltonian $H(N)$ of an open chain with $N$
sites. Then, the ground states of an open chain with $N+1$ lattice sites can
be written as the following superposition of $\left| \Psi _{\alpha ,\beta
}\right\rangle $ and $\left| m_{N+1}\right\rangle $:%
\begin{equation}
\left| \Psi _{N+1}\right\rangle =\sum_{\alpha \beta ,m_{N+1}}W_{\beta \alpha
,m_{N+1}}\left| \Psi _{\alpha ,\beta }\right\rangle \otimes \left|
m_{N+1}\right\rangle .
\end{equation}%
The vectors $\left| m_{N+1}\right\rangle $ on the site $N+1$ decouple from
the excited states of $H(N)$ because such a coupling do not gain energy from
the projector Hamiltonian $H(N+1)$. Now we can change the notation $W_{\beta
\alpha ,m_{N+1}}\equiv W_{\beta ,\alpha }^{[m_{N+1}]}$ and then $\left| \Psi
_{N+1}\right\rangle $ can be immediately written in a matrix-product form%
\begin{eqnarray}
\left| \Psi _{N+1}\right\rangle  &=&\sum_{m_{1}\cdots m_{N+1}}\mathrm{Tr}%
(A^{[m_{1}]}\cdots A^{[m_{N}]}W^{[m_{N+1}]})  \notag \\
&&\times \left| m_{1}\cdots m_{N+1}\right\rangle ,
\end{eqnarray}%
where $D\times D$ matrix $W^{[m_{N+1}]}$ can be fully determined because $%
\left| \Psi _{N+1}\right\rangle $ are the zero-energy ground states of $%
H(N+1)$. After solving this eigenvalue problem, one can find that $\left|
\Psi _{N+1}\right\rangle $ can be written in the form of Eq. (\ref{eq:MPSrep}%
). This final step completes the mathematical induction proof. For periodic
boundary conditions, the VBS ground state should be a linear combination of
the $D^{2}$ states in Eq. (\ref{eq:OBC-MPS}) and be annihilated by the extra
projectors acting on the two ends of the chain. This will lead to the VBS
ground state in the form of Eq. (\ref{eq:MPSrep}).

Actually, those matrix product states with $D^{2}$ linearly independent $%
\left| \Psi _{\alpha ,\beta }\right\rangle $ in a finite open chain satisfy
the so-called injective property. \cite{Fannes-1989,Cirac-2007,Sanz-2009}\
The injectivity of the matrix product states not only ensures the
ground-state uniqueness of the parent Hamiltonian, but also guarantees the
existence of an energy gap and the exponentially decaying correlation
functions of local operators.

\section{General VBS states for quantum integer spin chains}

In Sec. II, we set up a framework to study the VBS states with a built-in
Lie group $G$. To test these abstract formalism, we begin with the quantum
integer spin-$S$ chains and consider two classes of VBS states. In these two
VBS classes, the virtual particles transform under spin-$J$ representations
and $2^{S}$-dimensional $SO(2S+1)$ spinor representations, respectively.
Toward the first class, Sanz \textit{et. al.} \cite{Sanz-2009}\ have
explored $SU(2)$-invariant two-body Hamiltonians which have such states as
their eigenstates. In the present work, we further require that these VBS
states are unique ground states of the parent Hamiltonians. However, the
price we usually have to pay is to include multi-spin interactions in the
parent Hamiltonians. This situation will be treated for the spin-$1$
fermionic VBS state in Sec. III B. For the second class, we will show that
these states are equivalent to the $SO(2S+1)$ symmetric matrix product
states introduced in our previous work. \cite{Tu-2008} However, the present
formalism explains the origin of the emergent $SO(2S+1)$ symmetry in these
VBS states and shows that their edge states are $SO(2S+1)$ spinors.
Therefore, the two VBS classes are sharply distinct from each other for $%
S\geq 3$.

\subsection{Spin-$S$ VBS states with virtual spin-$J$ particles}

As a warm up, let us apply the formalism in Sec. II to the spin-$S$ VBS
states with two virtual spin-$J$ particles in each site. It is well-known
that the product of two spin-$J$ representation%
\begin{equation}
J\otimes J=0\oplus 1\oplus \cdots \oplus 2J
\end{equation}%
always contains a singlet and the physical spin-$S$ representation if $J\geq
S/2$. After replacing the $SU(2)$ Clebsch-Gordan coefficient $P_{\alpha
,\beta }^{[m]}=\langle S,m|J,\alpha ;J,\beta \rangle $ in Eq. (\ref%
{eq:projection}) and the spin-$J$ valence-bond singlet%
\begin{equation}
\left\vert I\right\rangle _{ij}=\sum_{\alpha =-J}^{J}(-1)^{J-\alpha
}\left\vert \alpha \right\rangle _{i}\otimes \left\vert -\alpha
\right\rangle _{j},
\end{equation}%
in Eq. (\ref{eq:singlet}), the $(2J+1)\times (2J+1)$ matrix $A^{[m]}$ in Eq.
(\ref{eq:MPSrep}) can be written as
\begin{equation}
A^{[m]}=\sum_{\alpha ,\beta }(-1)^{J+\beta }\langle S,m|J,\alpha ;J,-\beta
\rangle \left\vert J,\alpha \right\rangle \left\langle J,\beta \right\vert ,
\end{equation}%
where $-J\leq \alpha ,\beta \leq J$. These $A^{[m]}$ matrices are rank $S$\
irreducible spherical tensors and satisfy the following commutation
relations:%
\begin{eqnarray}
\lbrack J_{z},A^{[m]}] &=&mA^{[m]},  \notag \\
\lbrack J_{\pm },A^{[m]}] &=&\sqrt{(S\mp m)(S\pm m+1)}A^{[m\pm 1]},
\end{eqnarray}%
where $J_{\pm }$ and $J_{z}$ generate the spin-$J$ representation of the $%
SU(2)$ algebra,%
\begin{eqnarray}
J_{\pm } &=&\sum_{\alpha }\sqrt{(J\mp \alpha )(J\pm \alpha +1)}\left\vert
J,\alpha \pm 1\right\rangle \left\langle J,\alpha \right\vert ,  \notag \\
J_{z} &=&\sum_{\alpha }\alpha \left\vert J,\alpha \right\rangle \left\langle
J,\alpha \right\vert .
\end{eqnarray}

For the celebrated VBS states of the AKLT models, i.e., the case of virtual
spin $J=S/2$, we can also use the Schwinger boson representation to express
the VBS states. In the Schwinger boson language, the spin operators are
expressed by
\begin{equation}
S_{i}^{+}=a_{i}^{\dagger }b_{i},S_{i}^{-}=b_{i}^{\dagger
}a_{i},S_{i}^{z}=(a_{i}^{\dagger }a_{i}-b_{i}^{\dagger }b_{i})/2,
\end{equation}
with a local constraint $a_{i}^{\dagger }a_{i}+b_{i}^{\dagger }b_{i}=2S$.
Then, the integer spin-$S$ VBS states of the AKLT models in a periodic chain
are expressed as \cite{Arovas-1988}
\begin{equation}
\left\vert \mathrm{AKLT}\right\rangle =\prod_{i}(a_{i}^{\dagger
}b_{i+1}^{\dagger }-b_{i}^{\dagger }a_{i+1}^{\dagger })^{S}\left\vert
\mathrm{v}\right\rangle ,  \label{eq:AKLT}
\end{equation}%
where $\left\vert \mathrm{v}\right\rangle $ is the vacuum with no particle
occupation. The matrix product form of these VBS states are obtained by
Totsuka and Suzuki \cite{Suzuki-1995}. Since the null space of two
neighboring sites is the total bond spin $S+1,\ldots ,2S$ channels, the VBS
states in Eq. (\ref{eq:AKLT}) are exact ground states of AKLT Hamiltonians %
\cite{Affleck-1987,Arovas-1988}%
\begin{equation}
H_{\text{\textrm{AKLT}}}=\sum_{i}%
\sum_{S_{T}=S+1}^{2S}J_{S_{T}}P_{S_{T}}(i,i+1),
\end{equation}%
where all $J_{S_{T}}>0$ and $P_{S_{T}}(i,j)$ is the projection operator on
total bond\ spin channel $S_{T}$. These $SU(2)$ invariant projection
operators can be written as polynomials of spin-exchange interactions $%
\mathbf{S}_{i}\cdot \mathbf{S}_{j}$ up to $2S$ powers%
\begin{equation}
P_{S_{T}}(i,j)=\prod\limits_{\substack{ S^{\prime }=0,  \\ S^{\prime }\neq
S_{T}}}^{2S}\frac{2\mathbf{S}_{i}\cdot \mathbf{S}_{j}+2S(S+1)-S^{\prime
}(S^{\prime }+1)}{S_{T}(S_{T}+1)-S^{\prime }(S^{\prime }+1)}.
\label{eq:projector}
\end{equation}

For $S/2<J<S$ cases, at first glance, the null space of two neighboring
sites is given by the total bond spin channels $2J+1,\ldots ,2S$, which is
smaller than the VBS states of AKLT model. According to Sec. II B, one may
conclude that\ next-nearest neighbor interactions are needed to be construct
their parent Hamiltonians. However, there is an exception: the VBS states
with $S=2$ and $J=3/2$. We will discuss this special case in Sec. III C.

\subsection{Spin-$1$ fermionic VBS states}

Now we consider $S=1$ VBS state with virtual spin $J=1$, which belongs to
the class in Sec. III A. Since $S=1$ is the only antisymmetric product of
two virtual $J=1$ particles, one can use the fermionic statistics to
implement the projection onto the physical $S=1$ subspace. Thus, the
physical $S=1$ states are written as%
\begin{equation}
\left\vert 1\right\rangle =c_{1}^{\dagger }c_{0}^{\dagger }\left\vert
\mathrm{v}\right\rangle ,\text{ }\left\vert 0\right\rangle =c_{1}^{\dagger
}c_{-1}^{\dagger }\left\vert \mathrm{v}\right\rangle ,\text{ }\left\vert
-1\right\rangle =c_{0}^{\dagger }c_{-1}^{\dagger }\left\vert \mathrm{v}%
\right\rangle .
\end{equation}%
The $SU(2)$ spin operators are $S_{i}^{a}=\sum_{\mu ,\nu =1}^{3}c_{i\mu
}^{\dag }S_{\mu \nu }^{a}c_{i\nu }$ $(a=x,y,z)$, where $S^{a}$ are the usual
$3\times 3$ spin-$1$ matrices. The total spin $\mathbf{S}_{i}^{2}=2$ on each
site is imposed by a local constraint $\sum_{\mu =1}^{3}c_{i\mu }^{\dag
}c_{i\mu }=2$.

In terms of these fermionic variables, the $S=1$ VBS state with virtual spin
$J=1$ can be exactly written as%
\begin{equation}
\left\vert \Psi _{1}\right\rangle =\prod\limits_{i=1}^{N}(c_{i,1}^{\dagger
}c_{i+1,-1}^{\dagger }-c_{i,0}^{\dagger }c_{i+1,0}^{\dagger
}+c_{i,-1}^{\dagger }c_{i+1,1}^{\dagger })\left\vert \mathrm{v}\right\rangle
,
\end{equation}%
which has a matrix product form in Eq. (\ref{eq:MPSrep}) with%
\begin{eqnarray}
A^{[1]} &=&%
\begin{pmatrix}
0 & -1 & 0 \\
0 & 0 & -1 \\
0 & 0 & 0%
\end{pmatrix}%
,A^{[0]}=%
\begin{pmatrix}
1 & 0 & 0 \\
0 & 0 & 0 \\
0 & 0 & -1%
\end{pmatrix}%
,  \notag \\
A^{[-1]} &=&%
\begin{pmatrix}
0 & 0 & 0 \\
1 & 0 & 0 \\
0 & 1 & 0%
\end{pmatrix}%
.  \label{eq:spin1-matrix}
\end{eqnarray}

Following the method in Sec. II B, we can construct the parent Hamiltonian
for this fermionic VBS state. To locate the null space, it is sufficient to
block three successive spins. The tensor decomposition of three spin-$1$
representation yields%
\begin{eqnarray}
1\otimes 1\otimes 1 &=&(0\oplus 1\oplus 2)\otimes 1  \notag \\
&=&1\oplus 0\oplus 1^{\prime }\oplus 2\oplus 1^{\prime \prime }\oplus
2^{\prime }\oplus 3,
\end{eqnarray}%
which provides a natural basis for block spins. In this basis, the block
states can be denoted by $\left| S_{12};S,M\right\rangle $, where $S$ and $M$
are total spin and magnetic quantum number of the three sites, $S_{12}$ is
the total spin of the first two sites. For the representations $1^{\prime }$
and $2$, we have $S_{12}=1$. For the representations $1^{\prime \prime }$
and $2^{\prime }$, $S_{12}=2$. For the representations $0$ and $3$, the
index $S_{12}$ can be suppressed and does not lead to misunderstanding.

After blocking the three spins, we obtain%
\begin{eqnarray}
g_{1}g_{2}g_{3}
&=&\sum_{m_{1}m_{2}m_{3}}A^{[m_{1}]}A^{[m_{2}]}A^{[m_{3}]}\left|
m_{1},m_{2},m_{3}\right\rangle   \notag \\
&=&\sum_{S,M}\sum_{S_{12}}C_{S_{12}}^{[S,M]}\left| S_{12};S,M\right\rangle ,
\label{eq:3siteblock}
\end{eqnarray}%
where the $3\times 3$ matrices $C_{S_{12}}^{[S,M]}$ are given by%
\begin{eqnarray*}
C_{S_{12}}^{[S,M]} &=&\sum_{m_{1}m_{2}m_{3}}\left\langle
S_{12},m_{1}+m_{2}|1,m_{1};1,m_{2}\right\rangle  \\
&&\times \left\langle S,M|S_{12},m_{1}+m_{2};1,m_{3}\right\rangle
A^{[m_{1}]}A^{[m_{2}]}A^{[m_{3}]}.
\end{eqnarray*}%
The matrices $C_{S_{12}}^{[S,M]}$ can be calculated by using Eq. (\ref%
{eq:spin1-matrix}). Firstly, we find that $C^{[0,0]}\neq 0$ and $C^{[3,M]}=0$%
. This means that the spin-$0$ singlet state is contained but the spin-$3$
states are absent in every three-site block. The other \textit{nonvanishing}
matrices $C_{S_{12}}^{[S,M]}$ satisfy the following equations:
\begin{eqnarray}
\text{ }C_{2}^{[2,M]} &=&\sqrt{3}C_{1}^{[2,M]},  \notag \\
C_{2}^{[1,M]} &=&-\sqrt{\frac{5}{3}}C_{1}^{[1,M]}=\frac{\sqrt{5}}{4}%
C_{0}^{[1,M]}.
\end{eqnarray}%
According to Eq. (\ref{eq:3siteblock}), the unnormalized states contained in
the $3$-site block $g_{1}g_{2}g_{3}$ are one spin-$0$ state, three spin-$1$
states
\begin{equation*}
4\left| 0;1,M\right\rangle -\sqrt{3}\left| 1;1,M\right\rangle +\sqrt{5}%
\left| 2;1,M\right\rangle
\end{equation*}%
with $-1\leq M\leq 1$, and five spin-$2$ states
\begin{equation*}
\left| 1;2,M\right\rangle +\sqrt{3}\left| 2;2,M\right\rangle
\end{equation*}%
with $-2\leq M\leq 2$. By using the Gram-Schmidt orthogonalization method,
we find that seven spin-$3$ states $\left| 3,M\right\rangle $ with $-3\leq
M\leq 3$, five spin-$2$ states
\begin{equation*}
\left| \phi _{2,M}\right\rangle =\sqrt{3}\left| 1;2,M\right\rangle -\left|
2;2,M\right\rangle
\end{equation*}%
with $-2\leq M\leq 2$, six spin-$1$ states
\begin{eqnarray*}
\left| \phi _{1,M}\right\rangle  &=&\sqrt{3}\left| 0;1,M\right\rangle
+4\left| 1;1,M\right\rangle , \\
\left| \varphi _{1,M}\right\rangle  &=&4\left| 0;1,M\right\rangle -\sqrt{3}%
\left| 1;1,M\right\rangle -\frac{19}{\sqrt{5}}\left| 2;1,M\right\rangle
\end{eqnarray*}%
with $-1\leq M\leq 1$ span the null space in the reduced density matrix of
the $3$-site block. Therefore, the $3$-site projector Hamiltonian for which
the spin-$1$ fermionic VBS state is the zero energy ground state is thus
given by%
\begin{eqnarray}
h &=&\lambda _{3}\sum_{|M|\leq 3}\left| 3,M\right\rangle \left\langle
3,M\right| +\lambda _{2}\sum_{|M|\leq 2}\left| \phi _{2,M}\right\rangle
\left\langle \phi _{2,M}\right|   \notag \\
&&+\sum_{|M|\leq 1}(\lambda _{1}\left| \phi _{1,M}\right\rangle \left\langle
\phi _{1,M}\right| +\lambda _{1}^{\prime }\left| \varphi _{1,M}\right\rangle
\left\langle \varphi _{1,M}\right| ),
\end{eqnarray}%
where all $\lambda _{3},\lambda _{2},\lambda _{1},\lambda _{1}^{\prime }>0$.

\begin{figure}[tbp]
\centering \includegraphics[scale=0.8]{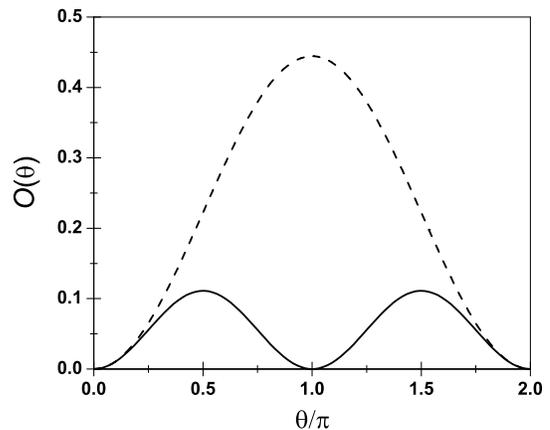}
\caption{The non-local string order parameter $\mathcal{O}(\protect\theta )$
as a function of the spin-twist angle $\protect\theta $ for the fermionic
VBS state (solid line) and the bosonic VBS state (dashed line) of the AKLT
model.}
\label{Fig:string}
\end{figure}

It is interesting to compare the spin-$1$ fermionic VBS state with the spin-$%
1$ bosonic VBS state of AKLT model. In the fermionic VBS state, the
two-point spin correlation function decays exponentially with a correlation
length $\xi =1/\ln 2$, longer than that for the AKLT model ($\xi =1/\ln 3$).
Besides the obvious difference of the edge spin representation in an open
chain, we can also see a sharp difference by computing the non-local string
order parameter \cite{Oshikawa-1992,Suzuki-1995}%
\begin{equation}
\mathcal{O}(\theta )=\lim_{\left\vert j-i\right\vert \rightarrow \infty
}\langle S_{i}^{z}\prod_{r=i}^{j-1}\exp (i\theta S_{r}^{z})S_{j}^{z}\rangle .
\label{eq:StringOrder}
\end{equation}%
By using the transfer matrix method \cite{Klumper-1991}, we arrive at the
result $\mathcal{O}(\theta )=\frac{1}{9}\sin ^{2}\theta $ for the fermionic
VBS state. For comparison, the values of the non-local order parameter $%
\mathcal{O}(\theta )$\ for both the spin-$1$ fermionic VBS state and bosonic
VBS state of the AKLT model are plotted in Fig. (\ref{Fig:string}). For the
spin-$1$ bosonic VBS state, $\mathcal{O}(\theta )$ reaches its maximum at $%
\theta =\pi $, which is reduced to the den Nijs-Rommelse string order
parameter characterizing the hidden antiferromagnetic order in the AKLT VBS
state.\ However, in the fermionic VBS state, $\mathcal{O}(\theta )$ has a
minimum $\mathcal{O}(\pi )=0$ and the maximum value $1/9$ at both $\theta
=\pi /2$ and $3\pi /2$. This signifies that the hidden antiferromagnetic
picture totally breaks down in the $S=1$ fermionic VBS state. How to
describe such a state has not been clear so far.

\subsection{VBS states with an emergent $SO(2S+1)$ symmetry}

In Sec. III A, we have mentioned an exceptional example: $S=2$ VBS states
with virtual spin $J=3/2$. Besides the absent bond total spin $S_{T}=4$
channel of neighboring sites, there is a new forbidden channel $S_{T}=2$ in
this VBS state. \cite{Tu-2008} Therefore, its parent two-body Hamiltonian
can be written as%
\begin{equation}
H=\sum_{i}\left[ J_{1}P_{2}(i,i+1)+J_{2}P_{4}(i,i+1)\right] ,
\label{eq:SO(5)AKLT}
\end{equation}%
with $J_{1},J_{2}>0$. According to Eq. (\ref{eq:projector}), the projector
Hamiltonian (\ref{eq:SO(5)AKLT}) can be rewritten as%
\begin{eqnarray}
H &=&\sum_{\left\langle ij\right\rangle }\left[ \frac{3J_{2}-80J_{1}}{84}%
\mathbf{S}_{i}\cdot \mathbf{S}_{j}+\frac{9J_{2}-40J_{1}}{360}(\mathbf{S}%
_{i}\cdot \mathbf{S}_{j})^{2}\right.  \notag \\
&&\left. +\frac{10J_{1}+J_{2}}{60}(\mathbf{S}_{i}\cdot \mathbf{S}_{j})^{3}+%
\frac{20J_{1}+J_{2}}{2520}(\mathbf{S}_{i}\cdot \mathbf{S}_{j})^{4}\right] .
\end{eqnarray}%
In fact, the $S=2$ VBS state with virtual spin $J=3/2$ has a hidden $SO(5)$
symmetry and its matrix-product form was studied by Scalapino \textit{et al}%
. \cite{Scalapino-1998}\ in an $SO(5)$ symmetric ladder system of
interacting electrons.

The quantum spin-$2$ Hamiltonian in Eq. (\ref{eq:SO(5)AKLT}) belongs to a
new class of exactly solvable quantum integer-spin chains introduced by the
first three of us very recently. \cite{Tu-2008}\ The ground states of these
Hamiltonians are $SO(2S+1)$ symmetric matrix-product states and exhibit
hidden topological order. For $S=1$, the $SO(3)$ symmetric matrix product
state becomes the VBS state of spin-$1$ AKLT model. For $S=2$, the $SO(5)$
symmetric matrix product state is the $S=2$ VBS state with virtual spin $%
J=3/2$. However, it was not clear whether this family of matrix product
states has a valence-bond picture for $S\geq 3$. By using the framework in
Sec. II, we will show that there is indeed\ a VBS picture for these matrix
product states. Actually, the virtual particles in these VBS states
transform under the $2^{S}$-dimensional spinor representation of $SO(2S+1)$.

It is convenient to promote the symmetry of the system and demand the spin-$%
S $ states on each site transform under the $(2S+1)$-dimensional vector
representation of $SO(2S+1)$. The tensor product of two $SO(2S+1)$ vectors
on the adjacent sites can be decomposed as
\begin{equation}
\underline{2S+1}\otimes \underline{2S+1}=\underline{1}\oplus \underline{%
S(2S+1)}\oplus \underline{S(2S+3)},  \label{eq:VectorDecom}
\end{equation}%
where the number above each underline is the dimension of the corresponding
IR. These $SO(2S+1)$ IRs can be directly related to $SU(2)$ integer-spin
IRs. Here $\underline{1}$ is the symmetric spin singlet, while the
antisymmetric channel $\underline{S(2S+1)}$ and the symmetric channel $%
\underline{S(2S+3)}$ correspond to the total bond spin $S_{T}=1,3,\ldots
,2S-1$ and $S_{T}=2,4,\ldots ,2S$ states, respectively. Therefore, the $%
SO(2S+1)$ bond projection operators can be expressed using the spin
projection operators as
\begin{eqnarray}
P_{\underline{S(2S+1)}}(i,j) &=&\sum_{l=1}^{S}P_{S_{T}=2l-1}(i,j), \\
P_{\underline{S(2S+3)}}(i,j) &=&\sum_{l=1}^{S}P_{S_{T}=2l}(i,j).
\end{eqnarray}

On each lattice site, the $SO(2S+1)$ vectors can be formed by tensor
decomposition of two virtual $2^{S}$-dimensional spinors%
\begin{equation}
\underline{2^{S}}\otimes \underline{2^{S}}=\bigoplus_{q=0}^{S}\underline{%
\binom{q}{2S+1}},  \label{eq:SpinorDecomp}
\end{equation}%
where $\binom{q}{2S+1}=\frac{(2S+1)!}{q!(2S-q+1)!}$. Note that $q=0$ and $%
q=1 $ in Eq. (\ref{eq:SpinorDecomp}) correspond to singlet representation
and $(2S+1)$-dimensional vector representation, respectively. For $S=1$, Eq.
(\ref{eq:SpinorDecomp}) recovers the well-known decomposition $\underline{2}%
\otimes \underline{2}=\underline{1}\oplus \underline{3}$ of two spin-$1/2$
spinors. For $S=2$, Eq. (\ref{eq:SpinorDecomp}) can be interpreted as the
decomposition $\underline{4}\otimes \underline{4}=\underline{1}\oplus
\underline{5}$ $\oplus \underline{10}$, where the $SO(5)$ spinors can be
viewed as spin-$3/2$ variables because $SO(5)\simeq Sp(4)$. However, the $%
SO(2S+1)$ spinors in Eq. (\ref{eq:SpinorDecomp}) for $S\geq 3$\ \textit{do
not} have $SU(2)$ spin counterparts.

Following the discussions in Sec. II, the $SO(2S+1)$ symmetric VBS states
can be constructed by combining the virtual spinors on the neighboring sites
into valence-bond singlets. By comparing Eq. (\ref{eq:VectorDecom}) and Eq. (%
\ref{eq:SpinorDecomp}), one finds that the IR channel $\underline{S(2S+3)}$
for any two neighboring sites\ is absent in these VBS states. Here an
interesting observation is that those IR channels with\ $q\geq 2$ in Eq. (%
\ref{eq:SpinorDecomp}) are actually absent for two adjacent sites due to the
projection of two virtual spinors onto the physical vector representation in
each site. Therefore, the $SO(2S+1)$-invariant parent Hamiltonian for the $%
SO(2S+1)$ symmetric VBS states is given by%
\begin{equation}
H=\sum_{i}P_{\underline{S(2S+3)}}(i,i+1).
\end{equation}%
Since the null space of these VBS states is the non-zero even total spin
channels, we can extend the $SO(2S+1)$-invariant parent Hamiltonian to the
following $SU(2)$-invariant quantum integer-spin Hamiltonian
\begin{equation}
H=\sum_{i}\sum_{l=1}^{S}J_{l}P_{S_{T}=2l}(i,i+1),
\end{equation}%
with all $J_{l}>0$.

Actually, the $SO(2S+1)$ symmetric VBS states are equivalent to the matrix
product states studied in Ref. \cite{Tu-2008}. In the present VBS form, the
origin of emergent $SO(2S+1)$ symmetry and the $2^{S}$ edge states on each
boundary of an open chain are quite clear. Although the edge degrees of
freedom in $S=1$ and $S=2$ cases can be viewed as $SU(2)$ spin variables,
they transform under $SO(2S+1)$\ spinor representation for $S\geq 3$ cases.
It is interesting to compare these $SO(2S+1)$ symmetric VBS states to the
spin-$S$ VBS states formed by virtual spin $J=(2^{S}-1)/2$ in Sec. III A.
Although they are both unique in a periodic chain and $4^{S}$ fold
degenerate in an open chain, their distinct edge states show that they
belong to two different topological classes. These explicit examples imply
that the ground state degeneracy is not sufficient to characterize the
topological ordered states.

\section{$SO(5)$ symmetric VBS states}

So far, we are restricted to the case of $SU(2)$ integer-spin in each site.
Actually, the method discussed in Sec. II can be applied for a general Lie
group $G$, we thus move on to $SO(5)$ Lie group, where the physical states
transform under $SO(5)$ IRs.

The $SO(5)$ Lie algebra has $10$ generators $L^{ab}$ $(1\leq a<b\leq 5)$,
satisfying the commutation relations%
\begin{equation}
\lbrack L^{ab},L^{cd}]=i(\delta _{ad}L^{bc}+\delta _{bc}L^{ad}-\delta
_{ac}L^{bd}-\delta _{bd}L^{ac}).  \label{eq:SO(5)CR}
\end{equation}%
Mathematically, the IRs of $SO(5)$ are labeled by two integers $(p,q)$, with
$p\geq q\geq 0$. For the $(p,q)$ representation of $SO(5)$ Lie group, the
dimensionality $d(p,q)$ and the Casimir charge $C(p,q)$ are given by \cite%
{SCZhang-2001}%
\begin{eqnarray}
d(p,q) &=&(1+q)(1+p-q)(1+\frac{p}{2})(1+\frac{p+q}{3}), \\
C(p,q) &=&\sum_{a<b}(L^{ab})^{2}=\frac{p^{2}}{2}+\frac{q^{2}}{2}+2p+q,
\label{eq:SO(5)Casimir}
\end{eqnarray}%
respectively. The dimensionality and Casimir charge for the simplest $SO(5)$
irreducible representations are listed in Tab. \ref{tab:IRs}.
\begin{table}[tbp]
\caption{Several irreducible representations of the $SO(5)$ Lie group.}
\label{tab:IRs}%
\begin{ruledtabular}
\begin{tabular}{lcr}
Representation & Dimension & Casimir charge\\
\hline
$(0,0)$   & 1 & 0\\
$(1,0)$   & 4 & 5/2\\
$(1,1)$   & 5 & 4\\
$(2,0)$   & 10 & 6\\
$(2,2)$   & 14 & 10\\
$(3,1)$   & 35 & 12\\
$(4,0)$   & 35 & 16
\end{tabular}
\end{ruledtabular}
\end{table}

\subsection{Bosonic $SO(5)$ VBS states}

\begin{figure}[t]
\centering \includegraphics[scale=1]{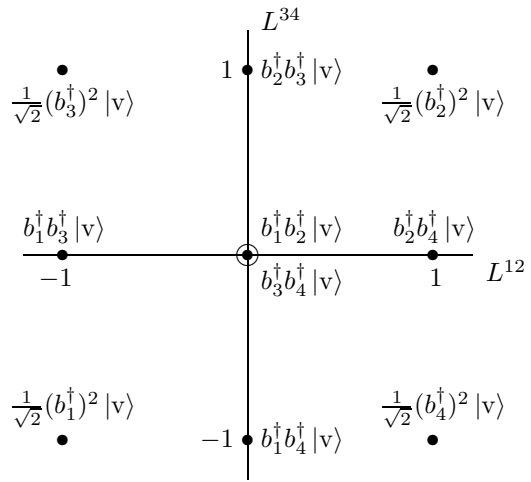}
\caption{Weight diagram and the bosonic realization of the $(2,0)$
adjoint
representation of $SO(5)$. There is a two-fold degeneracy with $%
L^{12}=L^{34}=0$.}
\label{Fig:Adjoint1}
\end{figure}

We begin with the $10$-dimensional $(2,0)$ adjoint representation of $SO(5)$%
. The bosonic $SO(5)/Sp(4)$ VBS state of this system was first considered by
Schuricht and Rachel. \cite{Rachel-2008} Their strategy is to construct the $%
(2,0)$\ adjoint representation by two virtual particles transforming under
the $(1,0)$\ spinor representation,%
\begin{equation}
(1,0)\otimes (1,0)=(0,0)\oplus (1,1)\oplus (2,0).  \label{eq:spinors}
\end{equation}%
where $(0,0)\ $and $(1,1)$ are antisymmetric and $(2,0)$ is the only
symmetric product representation. Therefore, one can obtain the physical $%
(2,0)$ adjoint representation by endowing bosonic statistics to the virtual $%
(1,0)$ spinor particles. This is analogous to the $SU(2)$ Schwinger boson
representation which symmetrizes two spin-$1/2$ spinors to construct a spin-$%
1$ representation. Using the $4$-component $SO(5)$ Schwinger\ bosons, the $%
SO(5)$ generators in Eq. (\ref{eq:SO(5)CR}) can be defined by%
\begin{equation}
L^{ab}=-\frac{1}{2}\sum_{\mu ,\nu =1}^{4}b_{\mu }^{\dag }\Gamma _{\mu \nu
}^{ab}b_{\nu },
\end{equation}%
where $\Gamma ^{ab}=[\Gamma ^{a},\Gamma ^{b}]/2i$ and%
\begin{equation}
\Gamma ^{1,2,3}=%
\begin{pmatrix}
0 & i\vec{\sigma} \\
-i\vec{\sigma} & 0%
\end{pmatrix}%
,\Gamma ^{4}=%
\begin{pmatrix}
0 & I \\
I & 0%
\end{pmatrix}%
,\Gamma ^{5}=%
\begin{pmatrix}
I & 0 \\
0 & -I%
\end{pmatrix}%
.
\end{equation}%
For the $(2,0)$ adjoint representation with $\sum_{\mu =1}^{4}b_{\mu }^{\dag
}b_{\mu }=2$, the $10$ states in a bosonic language are shown in the $(2,0)$
weight diagram in Fig. \ref{Fig:Adjoint1}. After a rotation by $45^{\circ }$%
, this weight diagram is identical to that given by Schuricht and Rachel.
Here we choose the Clifford algebra generated by the $\Gamma $\ matrices\ to
define the $SO(5)$ generators. The advantage of our convention is to find an
interesting non-local hidden string order in the $(2,0)$ bosonic VBS state
below.

The $(2,0)$ bosonic $SO(5)$ VBS state is formed by contracting two $(1,0)$
spinors on neighboring sites into a valence-bond $SO(5)$ singlet. Its wave
function can be written compactly as%
\begin{equation}
\left| \Psi _{2}\right\rangle =\prod_{i}(\sum_{\mu \nu }b_{i,\mu }^{\dag
}R_{\mu \nu }b_{i+1,\nu }^{\dagger })\left| \mathrm{v}\right\rangle ,
\end{equation}%
where the antisymmetric matrix $R$ is given by%
\begin{equation}
R=%
\begin{pmatrix}
-i\sigma ^{y} & 0 \\
0 & -i\sigma ^{y}%
\end{pmatrix}%
,
\end{equation}%
with the following properties:%
\begin{eqnarray}
R^{2} &=&-1,\text{ }R^{\dagger }=R^{-1}=R^{T}=-R,  \notag \\
R\Gamma ^{a}R^{-1} &=&(\Gamma ^{a})^{T},\text{ }R\Gamma ^{ab}R^{-1}=-(\Gamma
^{ab})^{T}.
\end{eqnarray}%
The tensor product decomposition of two neighboring $(2,0)$ adjoint
representations is written as%
\begin{equation}
(2,0)\otimes (2,0)=(0,0)\oplus (1,1)\oplus (2,0)\oplus (2,2)\oplus
(3,1)\oplus (4,0).
\end{equation}%
In the $(2,0)$ bosonic VBS state, a valence-bond singlet of two virtual $%
(1,0)$ spinors are created and therefore the two adjacent sites can only
transform $(0,0)$, $(1,1)$, and $(2,0)$ representations according to Eq. (%
\ref{eq:spinors}). Consequently, $\left| \Psi _{2}\right\rangle $ is an
exact ground state of the projector Hamiltonian%
\begin{eqnarray}
H &=&\sum_{i}\left[ J_{1}P_{(2,2)}(i,i+1)+J_{2}P_{(3,1)}(i,i+1)\right.
\notag \\
&&\left. +J_{3}P_{(4,0)}(i,i+1)\right] ,
\end{eqnarray}%
where $J_{1},J_{2},J_{3}>0$ and $P_{(2,2)},P_{(3,1)},P_{(4,0)}$ are
projectors onto the $(2,2),(3,1),(4,0)$ representations, respectively.

Furthermore, the $(2,0)$ bosonic VBS state contain a well-defined hidden
string order. This can be observed in its matrix-product wave function with
the local matrix%
\begin{equation*}
g_{i}=%
\begin{pmatrix}
\left| 0,0\right\rangle & \sqrt{2}\left| 1,1\right\rangle & \left|
0,1\right\rangle & \left| 1,0\right\rangle \\
-\sqrt{2}\left| -1,-1\right\rangle & -\left| 0,0\right\rangle & -\left|
-1,0\right\rangle & -\left| 0,-1\right\rangle \\
\left| 0,-1\right\rangle & \left| 1,0\right\rangle & \left| 0,0\right\rangle
^{\prime } & \sqrt{2}\left| 1,-1\right\rangle \\
-\left| -1,0\right\rangle & -\left| 0,1\right\rangle & -\sqrt{2}\left|
-1,1\right\rangle & -\left| 0,0\right\rangle ^{\prime }%
\end{pmatrix}%
_{i},
\end{equation*}%
where we take $\left| 0,0\right\rangle =b_{1}^{\dag }b_{2}^{\dagger }\left|
\mathrm{v}\right\rangle $ and $\left| 0,0\right\rangle ^{\prime
}=b_{3}^{\dag }b_{4}^{\dagger }\left| \mathrm{v}\right\rangle $. In both of
the $m_{1}$ and $m_{2}$ channels, it can be shown that $|m_{\eta }\rangle $ $%
(\eta =1,2)$ has a hidden antiferromagnetic order. In other word, the states
of $m_{\eta }=1$ and $-1$ will alternate in space if all the $m_{\eta }=0$
states between them are ignored. A typical configuration of this state is
given by
\begin{equation*}
\begin{array}{crcccccccccccccccl}
m_{1}: & \quad \cdots & 0 & \uparrow & 0 & 0 & \downarrow & \uparrow &
\downarrow & 0 & 0 & 0 & \uparrow & 0 & \downarrow & 0 & \uparrow & \cdots
\\
m_{2}: & \cdots & 0 & \uparrow & 0 & 0 & 0 & \downarrow & \uparrow & 0 &
\downarrow & 0 & 0 & \uparrow & \downarrow & 0 & 0 & \cdots%
\end{array}%
\end{equation*}%
where $(\uparrow ,0,\downarrow )$ represent $|m\rangle =(|1\rangle
,|0\rangle ,\left| -1\right\rangle )$. This hidden antiferromagnetic\ order
is in analogy with the spin-$1$ VBS state of AKLT model \cite{den Nijs-1989}%
\ and its $SO(2S+1)$ generalization \cite{Tu-2008}. To characterize this
hidden antiferromagnetic order, one can generalize the den Nijs-Rommelse
string order parameters as%
\begin{equation}
\mathcal{O}^{ab}=\lim_{|j-i|\rightarrow \infty }\langle
L_{i}^{ab}\prod_{r=i}^{j-1}\exp (i\pi L_{r}^{ab})L_{j}^{ab}\rangle .
\label{eq:SOP}
\end{equation}%
In fact, the non-local string order parameters for the Cartan generators
introduced by Schuricht and Rachel \cite{Rachel-2008} are combinations of
our $\mathcal{O}^{12}$ and $\mathcal{O}^{34}$. The advantage of our
convention is that the string order parameters in Eq. (\ref{eq:SOP}) clearly
reflect a hidden antiferromagnetic order in the $(2,0)$ bosonic $SO(5)$ VBS
state. The value of these string order parameter can be obtained by a
probability argument. These non-local string order parameters should all be
equal to each other because the VBS state preserves $SO(5)$ symmetry. Thus,
we only need to evaluate the value of $\mathcal{O}^{12}$ by considering $%
m_{1}$ channel. The role of the non-local string phase factor in Eq. (\ref%
{eq:SOP}) is to correlate the finite spin polarized states in the $m_{1}$
channel at the two ends of the string. If nonzero $m_{1}$ takes the same
value at the two ends, then the phase factor is equal to $1$. On the other
hand, if a nonzero $m_{1}$ takes two different values at the two ends, then
the phase factor is equal to $-1$. Thus, the value of $\mathcal{O}^{12}=9/25$
is a square of the probability of the nonzero $m_{1}=\pm 1$ appearing at the
ends of the string. Correspondingly, a generalized Kennedy-Tasaki unitary
transformation can be designed according to Ref. \cite{Tu-2008} and we
expect that the $SO(5)$ symmetry of the original Hamiltonian is reduced to $%
(Z_{2}\times Z_{2})^{2}$\ under such a non-local transformation. The
non-local string order parameters in Eq. (\ref{eq:SOP}) for the Cartan
generators will be transformed to two-point correlation functions, which
properly characterize the hidden $(Z_{2}\times Z_{2})^{2}$ symmetry
breaking. Thus, in this state, the non-local string order and the $16$-fold
degeneracy in an open chain can be viewed as natural consequences of a
hidden $(Z_{2}\times Z_{2})^{2}$ symmetry breaking.

In fact, the bosonic $SO(5)$ VBS state of $(2,0)$ adjoint representation can
be generalized to the totally symmetric $(p,0)$ representation with \textit{%
even} $p$. Namely, $(p,0)$ representation for even $p$ can be constructed by
two $(p/2,0)$ representations. However, there is an alternative way to take
the advantage of a generalized Schwinger boson representation. Using the
generalized Schwinger boson representation, the $(p,0)$ representation in
each site can be constructed by symmetrization of $p$ spinors and the local
constraint is now replaced with $\sum_{\mu =1}^{4}b_{\mu }^{\dag }b_{\mu }=p$%
. Thus, the bosonic $SO(5)$ VBS states for $(p,0)$ representation can be
written as%
\begin{equation}
\left| \Psi _{3}\right\rangle =\prod_{i}(\sum_{\mu \nu }b_{i,\mu }^{\dag
}R_{\mu \nu }b_{i+1,\nu }^{\dagger })^{p/2}\left| \mathrm{v}\right\rangle .
\label{eq:SO(5)VBS}
\end{equation}%
In an open chain, there are fractionalized edge states transforming under $%
(p/2,0)$ representation. Once a $\mathrm{CP}^{3}$ coherent state
representation \cite{Arovas-2008} is used, the $(p,0)$ VBS\ states have the
Jastrow form. They are analogous to the fractional quantum Hall states in
\textrm{CP}$^{3}$ space \cite{SCZhang-2003}\ at filling fraction $\nu =2/p$,
in the same sense as the resemblance \cite{Arovas-1988}\ between VBS states
of AKLT model and the fractional quantum Hall states in spherical geometry.

Since the tensor product decomposition of two $(p,0)$ representations is
given by%
\begin{equation}
(p,0)\otimes (p,0)=\sum_{k=0}^{p}\sum_{l=0}^{k}(k+l,k-l),
\label{eq:p0Decomp}
\end{equation}%
and $p/2$ valence-bond singlets are created between adjacent sites in $%
\left| \Psi _{3}\right\rangle $, the only finite projections on two adjacent
sites\ are given by
\begin{equation}
(p/2,0)\otimes (p/2,0)=\sum_{k=0}^{p/2}\sum_{l=0}^{k}(k+l,k-l).
\end{equation}%
Thus, the null space of the $2$-site\ reduced density matrix are given by a
sum of the representations written as $\sum_{k=p/2+1}^{p}%
\sum_{l=0}^{k}(k+l,k-l)$ and the corresponding parent Hamiltonian of $\left|
\Psi _{3}\right\rangle $ is given by%
\begin{equation}
H=\sum_{i}\sum_{k=\frac{p}{2}+1}^{p}%
\sum_{l=0}^{k}J_{(k+l,k-l)}P_{(k+l,k-l)}(i,i+1),
\end{equation}%
where all $J_{(k+l,k-l)}>0$ and $P_{(k+l,k-l)}$ is the projection operator
onto the $(k+l,k-l)$ representation states.

With the help of the Casimir charge in Eq. (\ref{eq:SO(5)Casimir}), the $%
SO(5)$ projectors can be written as polynomial functions of $SO(5)$
generators. According to Eq. (\ref{eq:p0Decomp}), these projectors satisfy a
completeness relation
\begin{equation}
\sum_{k=0}^{p}\sum_{l=0}^{k}P_{(k+l,k-l)}(i,j)=1.  \label{eq:completeness}
\end{equation}%
Considering the two-site Casimir charge $%
\sum_{a<b}(L_{i}^{ab}+L_{j}^{ab})^{2}$, we can write the $SO(5)$ Heisenberg
interaction as%
\begin{eqnarray}
\sum_{a<b}L_{i}^{ab}L_{j}^{ab} &=&\frac{1}{2}\sum_{k=0}^{p}%
\sum_{l=0}^{k}[C(k+l,k-l)-(p^{2}+4p)]  \notag \\
&&\times P_{(k+l,k-l)}(i,j).
\end{eqnarray}%
Using the properties of the projectors, we have%
\begin{eqnarray}
(\sum_{a<b}L_{i}^{ab}L_{j}^{ab})^{n} &=&\frac{1}{2}\sum_{k=0}^{p}%
\sum_{l=0}^{k}[C(k+l,k-l)-(p^{2}+4p)]^{n}  \notag \\
&&\times P_{(k+l,k-l)}(i,j).
\end{eqnarray}%
Together with the completeness relation (\ref{eq:completeness}), this
formula can be inverted, so that each projector can be represented by a
polynomial function of $SO(5)$ Heisenberg interaction $%
\sum_{a<b}L_{i}^{ab}L_{j}^{ab}$.

\subsection{Fermionic $SO(5)$ VBS state}

In this subsection, we present another way to construct the $(2,0)$ adjoint
representation, i.e., by using two $(1,1)$\ vector representations,
\begin{equation}
(1,1)\otimes (1,1)=(0,0)\oplus (2,0)\oplus (2,2),
\end{equation}%
where the $(2,0)$\ adjoint representation is antisymmetric and $(0,0),(2,2)$
are symmetric. This is because the orthogonal groups have a general property
that the adjoint representation is the only resulting antisymmetric channel
of two vector representations.\cite{Tu-2008} The simplest realization of
this property is the $SO(3)$ spin-$1$ case discussed in Sec. III\ B, where
the antisymmetrization of two vector spin-$1$\ representations only yields
the spin-$1$ adjoint representation.

If we use the fermionic statistics to implement the antisymmetrization, the $%
10$ states in the adjoint representation\ can be written as $c_{a}^{\dag
}c_{b}^{\dagger }\left| \mathrm{v}\right\rangle $, where $1\leq a,b\leq 5$.
Moreover, the $SO(5)$ generators are defined by%
\begin{equation}
L^{ab}=i(c_{a}^{\dag }c_{b}-c_{b}^{\dag }c_{a}),
\end{equation}%
and a double occupancy constraint $\sum_{a=1}^{5}c_{a}^{\dag }c_{a}=2$ can
guarantee the adjoint representation in each lattice site. Using these
fermionic variables, the $(2,0)$ weight diagram is shown in Fig. \ref%
{Fig:Adjoint2}.

\begin{figure}[t]
\centering \includegraphics[scale=1]{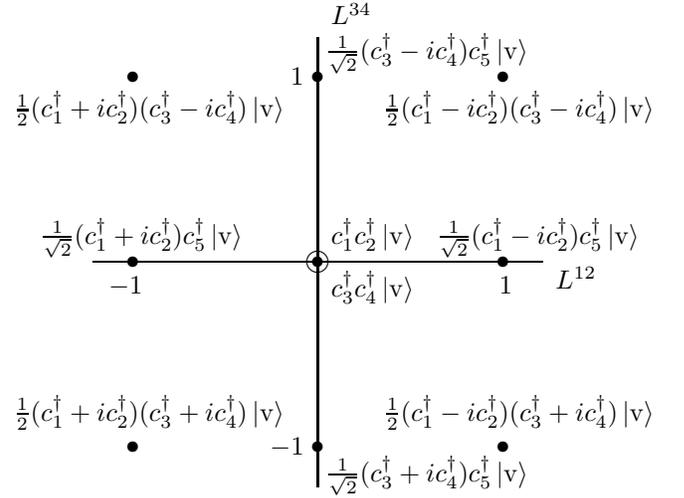}
\caption{Weight diagram and the fermionic realization of the
$(2,0)$ adjoint representation of $SO(5)$ Lie algebra.}
\label{Fig:Adjoint2}
\end{figure}

Using the fermion variables, the $(2,0)$ fermionic VBS state with two
virtual $(1,1)$ vector $SO(5)$ representations can be written as%
\begin{equation}
\left| \Psi _{4}\right\rangle =\prod_{i}(\sum_{a}c_{i,a}^{\dag
}c_{i+1,a}^{\dagger })\left| \mathrm{v}\right\rangle .
\end{equation}%
In an open chain, the edge spins transform under $(1,1)$ vector $SO(5)$
representation, different from the $(1,0)$ spinor $SO(5)$ representation in
the $(2,0)$ bosonic $SO(5)$ VBS state. Another interesting observation is
that the perfect non-local string order presence in the $(2,0)$ bosonic $%
SO(5)$ VBS state vanishes in the\ fermionic VBS state, because the string
order parameter (\ref{eq:SOP})\ for this state is found to be zero. In this
sense, the bosonic and fermionic $(2,0)$ VBS states can be viewed as $SO(5)$
generalizations of spin-$1$ VBS states of AKLT model and fermionic VBS state
in Sec. III\ B.

Finally, using two-body interactions, one can construct the parent
Hamiltonian for this fermionic $SO(5)$ VBS state. Since any two adjacent
sites can only transform under $(0,0)$, $(2,0)$, and $(2,2)$
representations, $\left| \Psi _{4}\right\rangle $ is an exact zero-energy
ground state of the projector Hamiltonian%
\begin{eqnarray}
H &=&\sum_{i}\left[ K_{1}P_{(1,1)}(i,i+1)+K_{2}P_{(3,1)}(i,i+1)\right.
\notag \\
&&\left. +K_{3}P_{(4,0)}(i,i+1)\right] ,
\end{eqnarray}%
for $K_{1},K_{2},K_{3}>0$. The possible hidden order is still under
investigation.

\section{Conclusion}

In conclusion, we have presented a general method to construct
one-dimensional VBS states embedded with Lie group $G$ and their parent
Hamiltonians. This provides examples that the topologically ordered states
can be systematically generated in one dimension and are characterized by
their edge states representations as well as their ground state degeneracy.

For quantum integer spin-$S$ chains, there exists two topologically distinct
families: (i) the virtual particles transform under $SU(2)$ spin-$J$
representations and (ii) the virtual particles are $SO(2S+1)$ spinors. In
the first class, a new spin-$1$ fermionic VBS state is constructed as an
explicit example. Compared to the celebrated $S=1$ valence bond solid state
of AKLT model, the fermionic valence bond solid state\ shows drastic
differences on the edge states and hidden string order. For the second
class, it has been shown that these valence bond solid states with an
emergent $SO(2S+1)$ symmetry are equivalent to the previously proposed $%
SO(2S+1)$ symmetric matrix-product states.\cite{Tu-2008} The present
formalism explicitly displays that the edge states of an open chain
transform under the $SO(2S+1)$ $2^{S}$-dimensional spinor representation.

To generalize the VBS states in $SU(2)$ symmetric quantum integer-spin
chains, two types of VBS states with the $SO(5)$ symmetry are considered,
including (i) bosonic $SO(5)$ VBS states formed by a symmetrization of two
spinor representations in each site and (ii) a fermionic $SO(5)$ VBS state
with $(2,0)$ adjoint representation formed by antisymmetrization of two
vector representations.

It can be expected that the ideas and formalism developed in this work are
very useful and can be generalized to the tensor product states (projected
entangled pair states) for higher dimensional correlated systems. \cite%
{jiang-xiang} The understanding of the physical properties of these states
is the first step to characterize higher dimensional topological states,
which certainly deserves further investigations.

\begin{acknowledgments}
One of the authors (H.H.Tu) would like to thank Mikel Sanz and J. Ignacio
Cirac for stimulating discussions during the visit in the Max Planck
Institute for Quantum Optics. He is also grateful to Stephan Rachel for
several helpful suggestions. We acknowledge the support of NSF-China and the
National Program for Basic Research of MOST-China.
\end{acknowledgments}

\end{document}